\documentclass[preprint]{aastex}
\usepackage{multicol,balance}
\usepackage{times}
\usepackage{graphicx}
\usepackage{subfigure}
\usepackage{amssymb}
\usepackage{txfonts}
\usepackage{}

\begin{document}

\title{Chromospheric activity on late-type star DM UMa using high resolution spectroscopic observations}

\author{LiYun Zhang\altaffilmark{1,2}, QingFeng Pi\altaffilmark{1,2},
Xianming L. Han \altaffilmark{1,3}, Liang Chang \altaffilmark{2,4}, and Daimei Wang$^{1,2}$}
\altaffiltext{1}{Department of Physics, College of Science, Guizhou
University and
NAOC-GZU-Sponsored Center for Astronomy, Guizhou
University, Guiyang 550025, P.R. China (e-mail: liy\_zhang@hotmail.com)}

\altaffiltext{2}{Key Laboratory for the Structure and Evolution of Celestial Objects of Yunnan Astronomical Observatories, Chinese Academy of Sciences, Kunming 650011, P.R. China}

\altaffiltext{3}{Dept. of Physics and Astronomy, Butler University, Indianapolis, IN 46208, USA}

\altaffiltext{4}{National
Astronomical Observatories/Yunnan Observatory, Chinese Academy of
Sciences, Kunming, 650011, P.R. China}
%



\begin{abstract}
We present new 14 high-resolution echelle spectra to discuss the level of chromospheric activity of DM UMa in $\mbox{He~{\sc i}}$ D$_{3}$, $\mbox{Na~{\sc i}}$\ D$_{1}$, D$_{2}$, H$_{\alpha}$, and $\mbox{Ca~{\sc ii}}$ infrared triplet lines (IRT). It is the first time to discover the emissions above the continuum in the $\mbox{He~{\sc i}}$ D$_{3}$ lines on Feb. 9 and 10, 2015. The emission on Feb. 9 is the strongest one ever detected for DM UMa. We analyzed these chromospheric active indicators by employing the spectral subtraction technique. The subtracted spectra reveal weak emissions in the $\mbox{Na~{\sc i}}$\ D$_{1}$, D$_{2}$ lines, strong emission in the H$_{\alpha}$ line, and clear excess emissions in the $\mbox{Ca~{\sc ii}}$ IRT lines. Our values for the $EW_{8542}$/$EW_{8498}$ ratio are on the low side, in the range of 1.0 - 1.7. There are also clear phase variations of the level of chromospheric activity in equivalent width ($EW$) light curves in these chromospheric active lines (especially the H$_{\alpha}$ line). These phenomena might be explained by flare events or rotational modulations of the level of chromospheric activity.
%

\end{abstract}


\section{Introduction}
\indent The prototype system RS CVn binary type was first proposed by Hall (1976), and later refined by Fekel et al. (1986) as the following: 1. A RS CVn binary system is a binary where at least one component shows strong $\mbox{Ca~{\sc ii}}$ H\&K emissions; 2. It displays light curve variations that can not be attributed to pulsation, eclipse or ellipticity; 3. The more active component is a F, G or K subgiant or giant, and it is evolved. These binary systems are characterized by strong levels of magnetic activity that present as photospheric starspots, chromospheric emissions, transition region emissions, coronal X-ray radiation, and flare events in optical, X-ray, ultraviolet, and radio wavelengths (Berdyugina 2005; Hall et al. 2008; Montes et al. 2004; G\"{u}del 2002, 2004; Zhang et al. 2015; etc). Studying RS CVn binary systems allows us to obtain important information concerning their magnetic activity, stellar dynamo, thus providing valuable constraints on theoretical models (Berdyugina 2005; Hall et al. 2008; etc).\\
\indent DM UMa (BD+61 1211, SAO 15338, K0-1 IV-III) is a single-lined spectroscopic binary with an orbital period of 7.5 days. It is one of the most active chromospheric active member of the RS CVn binary type (Crampton et al. 1979; Bopp 1982; Hatzes 1995; O'Neal et al. 1998; Eker et al. 2008; etc) and the optical counterpart of the X-ray source 2A 1052+606 (Liller et al. 1978; Charles et al. 1979; Schwart et al. 1979). Its level of magnetic activity has been observed from X-rays to radio wavelengths (Kashyap \& Drake 1999; Mutle \& Lestrade 1985; Wendker 1995; Mohin \& Raveendran 1992; Mitrou et al. 1996; Mitrou et al. 1997; Morris \& Mutel 1988).\\
\indent The H$_{\alpha}$ emission was first observed on two spectrograms by Liller et al. (1978). Later, Schwartz et al. (1979) found that the H$_{\alpha}$ emission was highly variable. Crampton et al. (1979) also confirmed the variability in H$_{\alpha}$ emission and observed for the first time the variable emissions in the $\mbox{Ca~{\sc ii}}$ H\&K lines. They also determined that the spectroscopic period of 7.492 days and the semi-amplitude of 28 $\pm$ 2 km~$s^{-1}$ from the radial velocity curve of DM UMa. Tan \& Liu (1985) presented five high-resolution observations, and all of them showed strong emission and variability in the H$_{\alpha}$ line. Nations \& Ramsey (1986) have reported that the strength of H$_{\alpha}$ emission in DM UMa varied by a factor of 3 over its orbital period during their 1991 observing run. They analyzed the excess $\mbox{Ca~{\sc ii}}$ H\&K, H$_{\epsilon}$ and H$_{\alpha}$ lines using the spectral subtraction method (Montes et al. 1995a; Montes et al. 1995b; Montes et al. 1996; Montes et al. 1997). The resulting subtraction spectra showed strong H$_{\alpha}$ emission with pronounced wings, emission in the $\mbox{Ca~{\sc ii}}$ H\&K lines, and weak emission in the H$_{\epsilon}$ line. Hatzes (1995) presented a doppler image of the spot distribution in 1993, and also found the behaviour of the chromospheric activity indicators of emission in the H$_{\alpha}$ line and absorption in the $\mbox{He~{\sc i}}$ D$_{3}$ line. Hatzes (1995) found that the $EW$ varied as a function of orbital phase in the H$_{\alpha}$ emission line and the $\mbox{He~{\sc i}}$ D$_{3}$ absorption line. \\
\indent Kimble et al. (1981) obtained the first light curve of DM UMa in 1979 and the photometric period of 7.4 days. They detected the light curve wave-like distortion due to the dark starspot with its amplitude of 0.36 mag in $B$ band and 0.32 mag in $V$ band. Mohin et al. (1985) obtained light curves in 1980 - 1984, and found that the shape and amplitude of the light curves had changed in the time scale of a few years, which they explained physically by using a two-starspot model. Heckert et al. (1988, 1990) and Heckert (1990) published the light curves of DM UMa in $UBV$ bands in 1986-1987, and 1988-1989. Mohin \& Raveendran (1994) determined that the variation amplitude increased by 0.2 mag in the $V$ magnitude, and that the spot position and area had changed during the period from 1988 to 1991. There was a phase dependence of spot filling factor vs. phase for DM Uma using a technique of molecular absorption bands (O'Neal 2004). Later, Rosario et al. (2009) presented extensive $UBVRI$ photometry at different observatories from 1988 to 2008. They obtained highly variable light curves and tried to search for evidence of any cyclic spot activities (Rosario et al. 2009). They discovered that the spot filling factors derived from the TiO-band strengths (O'Neal 2006) are anti-correlated with the $V$ magnitudes, and the excess flux in $U$ and $B$ bands originates from plages or facular regions associated with the spot activities (Rosario et al. 2009). Ta\c{s} and Evren (2012) discussed the multi-color photometric observations of DM UMa in the time interval between 1980 and 2009 (Heckert 2012; etc) and revealed the period estimations of 51.2 $\pm$ 2.8 years and 15.1 $\pm$ 0.7 years superimposed on it. This cyclic variation on a timescale of 17 year was found based on observations from a time span of twenty eight years, and was caused by temperature changes or hot surface regions (Ol\'{a}h et al. 2014).\\
\indent Chromospherically active stars and their properties have been studied previously by many authors (Strassmeier et al. 1990, 1993; Doyle et al. 1994; Fekel et al. 2005; Montes et al. 2004; Hall 2008; Biazzo et al. 2009; Frasca et al. 2010; Cao \& Gu 2015; Zhang et al. 2015; etc). The spectral subtraction technique has been widely applied to different chromospheric activity indicators (Barden 1985; Montes et al. 2004; Zhang et al. 2015; etc). In this paper, we present additional high-resolution spectroscopic observations of DM UMa, and the analyses of the behaviour and properties of the level of chromospheric activity in the $\mbox{He~{\sc i}}$ D$_{3}$, $\mbox{Na~{\sc i}}$\, D$_{1}$, D$_{2}$, H$_{\alpha}$, and $\mbox{Ca~{\sc ii}}$ IRT lines using the spectral subtraction technique. \\

\section{Observations}

We carried out new high-resolution spectroscopic observations of DM UMa on fourteen nights from Jan. 24 - Feb. 15, 2015 using the Coud\'{e} echelle spectrograph on the 1.8-m telescope (Wei et al. 2000; Rao et al. 2008) at Lijiang Station of the Yunnan Astronomical Observatories of China. A 2048$\times$2048 pixel Tektronix CCD detector is used and the spectral resolution is about 50, 000 in the red spectral region of 5760-11960 ${\AA}$. The limiting magnitude of the equipment is about 11.5 mag at present. The spectral resolution in terms of the FWHM of the arc comparison lines is 0.112 ${\AA}$ in the $\mbox{He~{\sc i}}$ D$_{3}$, $\mbox{Na~{\sc i}}$\, D$_{1}$, D$_{2}$ lines, 0.119 ${\AA}$ in the H$_{\alpha}$ line, 0.133 ${\AA}$ in the $\mbox{Ca~{\sc ii}}$ IRT 8498 ${\AA}$ line, 0.145 ${\AA}$ in the $\mbox{Ca~{\sc ii}}$ IRT 8542 ${\AA}$ line, and 0.150 ${\AA}$ in the $\mbox{Ca~{\sc ii}}$ IRT 8662 ${\AA}$ line, respectively. We observed several similar inactive stars (HR 617 (K1 III), $\beta$ Gem (K0 III), HR 4182 (G9 IV) and HR 495 (K2 IV)) whose spectral types and luminosity classes are similar to those of DM UMa, which are used to construct the synthesized spectra as the stellar photospheric contribution. We performed standard reduction of the spectra using IRAF packages\footnote {IRAF is distributed by the National Optical Astronomy Observatories (NOAO) in Tucson, Arizona, which is operated by the Association of Universities for Research in Astronomy, Inc., under cooperative agreement with the National Science Foundation.}, which involved CCD trim, bias subtraction, flat-field correction, cosmic-ray removal, background subtraction, and multi-spectrum extraction. We calibrated the wavelength using the spectra of a Th-Ar lamp. We normalized the observed spectra by a five-order polynomial fit. The normalized spectra are plotted in Fig. 1 (Left panel). We listed our observing log of DM UMa in Table 1, which included the observational data, the corresponding Heliocentric Julian Date (HJD), the observational exposure times, and the signal to noise ratios of different chromospheric active indicators.
\section{Analyses of the chromospheric activity}
We obtained a total of fourteen spectra. Among these spectra, the $\mbox{Na~{\sc i}}$\, D lines show deep absorption but no obvious self-reversal core emission. The H$_{\alpha}$ line in all the spectra exhibits clear emission above the continuum. We calculated the $EW$s and line depths of DM UMa for the $\mbox{Na~{\sc i}}$\, D$_{1}$, D$_{2}$, and H$_{\alpha}$ lines. The line depth is used as a diagnostic criterion for the level of chromospheric activity (Mallik 1997; Eaton 1995; Mallik 1998; etc). All the $\mbox{Ca~{\sc ii}}$ IRT lines show filled-in absorption with self-reversal core emission, with some of them being above the continuum. For the $\mbox{Ca~{\sc ii}}$ IRT lines, it is impossible to measure the $EW$ and line depth in the observed spectra because they exhibit clear self-reversal. The results are listed in Table 2, where the orbital phases of our observations were calculated using the ephemeris formula (HJD=2447623.383 +$ 7^{d}.4949$ E) given by Strassmeier et al. (1993). The $EW$s of the chromospheric active lines were evaluated on the spectra by integrating them over the emission profile using the IRAF {\it SPLOT} task. The methods for calculating the $EW$s and their uncertainties were similar to those in the previous paper in detail (Zhang \& Gu 2008; Zhang et al. 2015; etc). We plotted the observed $EW$s and line depths of DM UMa vs phase or HJD in Fig. 2, where different symbols represent different chromospheric active indicators. As can be seen from Fig. 2, there are clear phase and time variation in the $EW$s of the H$_{\alpha}$ line. These phenomenon might be explained by flare events or by chromospheric rotational modulations.\\
\indent It is well known that the $\mbox{He~{\sc i}}$ D$_{3}$ line is a probe for detecting flare-like events (Zirin 1988). Here we report that we detected a clear emission in the $\mbox{He~{\sc i}}$ D$_{3}$ line on Feb. 09 in 2015 and a weak emission on Feb. 10, 2015, which had never been detected previously. The $EW$s of the observed spectra in DM UMa are 0.118 $\pm$ 0.013 ${\AA}$ on Feb 9, 2015, and 0.042 $\pm$ 0.002 ${\AA}$ on Feb 10, 2015 respectively. For better comparisons, we plotted the spectra with $\mbox{He~{\sc i}}$ D$_{3}$ lines for three consecutive days in Fig. 3. For the spectrum on Feb. 8, there are no emission in the $\mbox{He~{\sc i}}$ D$_{3}$ line. The obvious emissions in the $\mbox{He~{\sc i}}$ D$_{3}$ line in the Feb. 9 \& 10 spectra means that there might be strong flare-like events in these two days.\\
\indent We analyzed our DM UMa spectra with a synthetical spectral subtraction technique using the {\it STARMOD} program developed by Barden (1985) and Montes et al. (1997). The synthesized spectra were built from radial-velocity shifted and rotationally broadened spectrum of a single inactive star. $\beta$ Gem was the best template star to construct such a synthesized spectrum. In the program, we determined the rotational velocity ($v$sin$i$) value of DM UMa using the 11th order spectrum in the wavelength range 6387-6487 {\AA} because there are a lot of independent metallic spectra lines (Zhang \& Gu 2008). Using the {\it STARMOD} program, our calculated $v$sin$i$ of DM UMa (28.7 km~$s^{-1}$) is close to the previous results of 26 km~$s^{-1}$ derived by Hatzes (1995), 27 km~$s^{-1}$ by Tan \& Liu (1985) and 28 km~$s^{-1}$ by Crampton et al. (1979). We plotted all the observed, synthesized and subtracted spectra of DM UMa in Fig. 1. We applied the spectral subtraction technique to these spectra and the results reveal that the cores of the $\mbox{Na~{\sc i}}$\, D$_{1}$, D$_{2}$ excess lines show weak emissions (see Fig. 1). All the observed spectra exhibit clear H$_{\alpha}$ emission above the continuum, and their corresponding excess spectra show broad and strong excess emissions. The behaviour of the H$_{\alpha}$ line is consistent with that obtained by Montes et al. (1995a, 1995b), Hatzes (1995), and Montes et al. (1997). The excess spectra of the $\mbox{Ca~{\sc ii}}$ IRT lines show clear excess emissions in their cores. The parameters of the excess emissions are listed in Table 3, which include HJD, the phase, the net $EW$s of $\mbox{Na~{\sc i}}$ D$_{1}$, D$_{2}$, H$_{\alpha}$ and $\mbox{Ca~{\sc ii}}$ IRT lines and the ratio of $\mbox{Ca~{\sc ii}}$ $EW_{8542}$ to $EW_{8498}$. We also plotted them in Fig. 2.\\
\indent The value of the ratio ($EW_{8542}/EW_{8498}$) is also an indicator for the chromospheric active phenomenon and was used to differentiate the plage or prominence. Our values for the ratio $EW_{8542}$/$EW_{8498}$ of DM UMa are about 1.0-1.7 (see Table 3). These small values mean that there are optically thick emissions in possible stellar plage regions. These low values were also discovered in many late-type stars (Ar\'{e}valo \& L\'{a}zaro 1999; Cao \& Gu 2012; Gu et al. 2002; G\'{a}lvez et al. 2009; Montes et al. 2001; Zhang et al. 2015).\\
\section{Discussions and conclusion}
\indent The DM UMa spectra that we obtained exhibit deep absorptions in the $\mbox{Na~{\sc i}}$\ D$_{1}$, D$_{2}$ lines, strong emissions above continuum in the H$_{\alpha}$ line, clear self-reversal emissions (some of them are above continuum) in the strong filled-in absorptions in the $\mbox{Ca~{\sc ii}}$ IRT lines. The corresponding subtraction spectra reveals weak emissions in the cores of the $\mbox{Na~{\sc i}}$\, D$_{1}$, D$_{2}$ lines, and strong emission in the H$_{\alpha}$ line and clear emissions in the $\mbox{Ca~{\sc ii}}$ IRT lines. The excess chromospheric emission is larger than the results obtained by Montes et al. (1995a, 1995b, 1997), and Hatzes (1995). It is the largest emission with an $EW$ of 5.586 ${\AA}$ for the observed spectra (see Table 2) and 7.839 ${\AA}$ (see Table 3) for the excess emission of DM UMa on Feb 9, 2015. All these confirmed the high level of chromospheric activity of DM UMa.\\
\indent Our data allowed us to determine various chromospheric properties of DM UMa. We found that there is phase or time variation in the level of chromospheric activity of the $\mbox{Na~{\sc i}}$ D$_{1}$, D$_{2}$, H$_{\alpha}$ and $\mbox{Ca~{\sc ii}}$ IRT lines (especially the H$_{\alpha}$ line). The orbital phases of all the $\mbox{Na~{\sc i}}$ D$_{1}$, D$_{2}$,  H$_{\alpha}$ and $\mbox{Ca~{\sc ii}}$ IRT lines with enhanced emissions were consistently located at around 0.5 (Fig. 2). The variations in the $EW$s of $\mbox{Na~{\sc i}}$\, D$_{1}$, D$_{2}$ and $\mbox{Ca~{\sc ii}}$ IRT are relative weak. This may be due to that they are the weak active lines among these chromospheric active indicators and affected by the presence of spot regions (Andretta et al. 1997; Montes et al. 1997). It is very difficult to distinguish flare events from cyclical variations for highly active stars. There is a slow rise and fall in the $EW$s of chromospheric active emission in the H$_{\alpha}$ line. {These $EW$ variations may be due to chromospheric rotational modulations, flare events or strong plages. In order to pinpoint which of these is the ultimate cause for this $EW$ variation, we need complete phase coverage in a short time, such as several consecutive nights.\\
\indent In order to study DM UMa's long term chromospheric variability, we compare in Fig. 4 our observed and chromospheric excess $EW$s with those provided by several authors.} As can be seen from Fig. 4, $EW$ does vary with the orbital phase. However, at different observation times, this phase dependence varies. The orbital phase of our chromospheric rotation modulation is similar to that obtained by Hatzes (1995) with the maximum occurring at a phase of 0.5, which is different from those obtained by Mohin \& Raveendran (1994) and Nation \& Ramsey (1986). The evidence of active longitudes in the chromosphere were also observed in other chromospherically active stars, such as PW And (L\'opez-Santiago et al. 2003; Zhang et al. 2015), V383 Lac (Biazzo et al. 2009; Zhang et al. 2015), VY Ari, IM Peg, HK lac (Biazzo et al. 2006), V889 Her (Frasca et al. 2010), LQ Hya (Cao \& Gu 2014; Zhang et al. 2014; Flores Soriano et al. 2015), V711 Tau (Garc\'{i}a--Alvarez et al. 2003; Cao \& Gu 2015), SZ Psc (Zhang \& Gu 2008; Cao \& Gu 2012), II Peg (Frasca et al. 2008; Gu et al. 2015), UX Ari (Gu et al. 2002), HD 22049 and HD 166 (Biazzo et al. 2007). The variation of the phase corresponding to maximum $EW$ for DM UMa from different observation times indicates that there are long-term variations in the level of chromospheric activity for DM UMa. The evolution of chromospheric activity was also found in several other chromospherically active stars, such as PW And, LQ Hya, SZ Psc and V711 Tau (L\'opez-Santiago et al. 2003; Cao \& Gu 2014; Cao \& Gu 2012; Garc\'{i}a--Alvarez et al. 2003).\\
\indent The correlation between photospheric starspots and chromospheric plages has been studied using the simultaneous photometric and high-resolution spectroscopic observations (Biazzo et al. 2006; L\'opez-Santiago et al. 2003; Frasca et al. 2010; Cao \& Gu 2014; Zhang et al. 2014; etc). Because there are no simultanenous photometric light curves with our high-resolution observations of DM UMa, we are not able to directly study the relationship between photospheric starspot activity and chromospheric activity using the method, as done by the above authors. Moreover, the starspot parameters could also be obtained using the ratio of the depth of atomic line (Catalano et al. 2002; O'Neal 2006) and molecular absorption bands (O'Neal et al. 1998; O'Neal 2004). The orbital phase of our chromospheric activity is similar to that of starspot in several light curves, and different with other light curves (Rosario et al. 2009; Ta\c{s} \& Evren 2012; Heckert 2012; O'Neal 2014). Hatzes (1995) found the maxima for H$_{\alpha}$ emission and $\mbox{He~{\sc i}}$ D$_{3}$ absorption was coincident in time with the starspot distribution by Doppler imaging. All these results gave supporting evidence for the existence of active longitudes of the level of magnetic activity in DM UMa. Although there are long-term variations in the level of chromospheric activity, more data are needed to search for a chromospheric active cycle, similar to that for the photometric cycle (Rosario et al. 2009; Ta\c{s} \& Evren 2012; Ol\'{a}h et al. 2014).\\
\indent The $\mbox{He~{\sc i}}$ D$_{3}$ (5876 ${\AA}$) is a valuable probe for the stellar activity. For the Sun, there is strong absorption in the $\mbox{He~{\sc i}}$ D$_{3}$ line for plages and weak flares (Landman 1981), and emission for the strong flares (Zirin 1988). Hatzes (1995) first discussed the absorption of $\mbox{He~{\sc i}}$ D$_{3}$ line in DM UMa and they also found that there were variable absorptions. Most of our DM UMa spectra have weak absorptions, with two $\mbox{He~{\sc i}}$ D$_{3}$ emission lines having above the continuum emissions during the observations, which may be caused by the flare events. The $EW$ excess for $\mbox{He~{\sc i}}$ D$_{3}$ emission is 0.166 $\pm$ 0.006 ${\AA}$ in Feb 9, 2015, and 0.066 $\pm$ 0.006 ${\AA}$ on Feb 10, 2015. The Feb 9, 2015 $\mbox{He~{\sc i}}$ D$_{3}$ emission is the strongest emission ever detected for DM UMa. As can be seen from Fig. 1, the wing of H$_{\alpha}$ line is broadening while the $\mbox{He~{\sc i}}$ D$_{3}$ line becomes an emission line. Moreover, the $EW$s are also increasing in the excess emission in the $\mbox{Na~{\sc i}}$ D$_{1}$, D$_{2}$, H$_{\alpha}$ and $\mbox{Ca~{\sc ii}}$ IRT lines on these days (Figs. 2). This behavior of the chromospheric activity indicators is consistent with a flare-like event. $\mbox{He~{\sc i}}$ D$_{3}$ emission serves as a very useful tool in studying stellar activity because it is found in other active stars, such as II Peg (Huenemoerder \& Ramsey 1987; Frasca et al. 2008; Gu \& Tan 2003; Berdyugina et al. 1999; Gu et al. 2015), V410 Tau (Welty \& Ramsey 1995), UX Ari (Montes et al. 1996b; Montes et al. 1997; Gu et al. 2002), HR 1099 (Garc\'{i}a-Alvarez et al. 2003), and LQ Hya (Montes et al. 1999). \\
\indent In order to compare the level of chromospheric activity between DM UMa and other stars, we collected the parameters and active properties of chromospheric active system with spectral type and rotational velocity similar to those for DM UMa. The results are shown in Table 4 (parameters are taken from the paper of Eker et al. (2008)), which includes star name, spectral type, magnitude, orbital period, rotational velocity, and the behaviour of the $\mbox{He~{\sc i}}$ D$_{3}$, $\mbox{Ca~{\sc ii}}$ H\&K and H$_{\alpha}$ lines. By comparing the behaviour of $\mbox{Ca~{\sc ii}}$ H\&K and H$_{\alpha}$ lines, the emission of $\mbox{He~{\sc i}}$ D$_{3}$ and the values of chromospheric $EW$s of H$_{\alpha}$ line, DM UMa has the highest level of chromospheric activity among these stars. It is also the only one with $\mbox{He~{\sc i}}$ D$_{3}$ emission. Therefore DM UMa is a very interesting target to study stellar magnetic activity, and warrants further investigations.\\
\acknowledgments{This work was supported in part by the Joint Fund of Astronomy of the National Natural Science Foundation of China (NSFC) and the Chinese Academy of Sciences (CAS) Nos. U1431114 and 11263001. We would like to thank reviewer's comments and suggestions, which led to great improvement in our manuscript. This work was also partially supported by the Open Project Program of the Key Laboratory of Optical Astronomy, NAOC, CAS. This research has made use of the SIMBAD database, operated at CDS, Strasbourg, France. We are very grateful to Dr. Gu and Dr. Montes for
their help. }

\begin{table*}
\small\tabcolsep 0.08cm \caption{Observational log of DM UMa.}
\begin{tabular}{cllcccccccc}
 \hline \multicolumn{1}{c}{Date} &
\multicolumn{1}{c}{HJD} & \multicolumn{1}{c}{Exp.
time} & \multicolumn{5}{c}{$S/N$} \\
 \hline  &2450000+ days & (s) & $\mbox{Na~{\sc
i}}$&H$_{\alpha}$& Ca II IRT$\lambda$8498&Ca II
IRT$\lambda$8542& Ca II IRT$\lambda$8662\\
\hline
 1/24/2015    &   7046.35136  &  1800   &   40   &     46     &   40    &     47   &     46 \\
 1/26/2015    &   7049.35378  &  2400   &   52   &     58     &   47    &     53   &     50 \\
 1/27/2015    &   7050.32868  &  2400   &   55   &     64     &   50    &     57   &     52 \\
 1/28/2015    &   7051.32824  &  2400   &   56   &     63     &   60    &     61   &     61 \\
 2/03/2015    &   7057.33445  &  3600   &   50   &     58     &   55    &     64   &     62 \\
 2/05/2015    &   7059.31556  &  3600   &   59   &     68     &   55    &     62   &     56 \\
 2/06/2015    &   7060.36109  &  2400   &   51   &     60     &   48    &     57   &     54 \\
 2/07/2015    &   7061.33654  &  3600   &   63   &     73     &   63    &     70   &     65 \\
 2/08/2015    &   7062.25021  &  2400   &   53   &     61     &   49    &     55   &     53 \\
 2/09/2015    &   7063.26832  &  3600   &   66   &     82     &   68    &     78   &     74 \\
 2/10/2015    &   7064.25727  &  3600   &   64   &     77     &   60    &     71   &     67 \\
 2/12/2015    &   7066.26788  &  3600   &   62   &     71     &   58    &     70   &     66 \\
 2/14/2015    &   7068.27980  &  3600   &   69   &     80     &   65    &     77   &     70 \\
 2/15/2015    &   7069.22819  &  3600   &   46   &     62     &   49    &     61   &     59 \\

\hline
\end{tabular}
\end{table*}

\begin{table*}
\tiny\tabcolsep 0.45cm \caption{Our $EW$ results and line depths for the chromospheric active indicators of the observed $\mbox{Na~{\sc i}}$\, D$_{1}$, D$_{2}$,
and H$_{\alpha}$ lines for DM UMa.}
\begin{tabular}{lccccccccccll}
\hline \multicolumn{1}{c}{HJD(2450000+)}
&\multicolumn{1}{c}{Phase}
&\multicolumn{2}{c}{EW$_{\rm H_{\alpha}}$(${\AA}$)}
&\multicolumn{2}{c}{EW$_{\rm Na_{I}D_{1}}$(${\AA}$)}
&\multicolumn{2}{c}{EW$_{\rm Na_{I}D_{2}}$(${\AA}$)}
\\
\hline
days &  -  &  $EW$   &   line depth  &  $EW$   &   line depth &  $EW$   &   line depth  \\
\hline
7046.35136 &  0.251  &  2.929  $\pm$  0.043    &    1.717 $\pm$   0.007   &  0.357  $\pm$  0.020 &   0.405 $\pm$   0.003  &   0.97   $\pm$ 0.122  &  0.796  $\pm$  0.074\\
7049.35378 &  0.651  &  1.909  $\pm$  0.009    &    1.517 $\pm$   0.002   &  0.368  $\pm$  0.019 &   0.401 $\pm$   0.012  &   1.066  $\pm$ 0.125  &  0.790  $\pm$  0.077\\
7050.32868 &  0.781  &  1.142  $\pm$  0.016    &    1.423 $\pm$   0.006   &  0.366  $\pm$  0.003 &   0.386 $\pm$   0.001  &   1.017  $\pm$ 0.069  &  0.796  $\pm$  0.001\\
7051.32824 &  0.915  &  1.223  $\pm$  0.009    &    1.479 $\pm$   0.008   &  0.341  $\pm$  0.011 &   0.390 $\pm$   0.001  &   1.057  $\pm$ 0.035  &  0.820  $\pm$  0.028\\
7057.33445 &  0.716  &  1.462  $\pm$  0.006    &    1.507 $\pm$   0.009   &  0.351  $\pm$  0.010 &   0.392 $\pm$   0.002  &   1.082  $\pm$ 0.133  &  0.777  $\pm$  0.062\\
7059.31556 &  0.980  &  1.211  $\pm$  0.007    &    1.448 $\pm$   0.009   &  0.359  $\pm$  0.005 &   0.382 $\pm$   0.017  &   0.998  $\pm$ 0.213  &  1.065  $\pm$  0.164\\
7060.36109 &  0.120  &  1.591  $\pm$  0.098    &    1.586 $\pm$   0.008   &  0.339  $\pm$  0.006 &   0.397 $\pm$   0.004  &   1.082  $\pm$ 0.068  &  0.811  $\pm$  0.010\\
7061.33654 &  0.250  &  1.485  $\pm$  0.055    &    1.571 $\pm$   0.009   &  0.374  $\pm$  0.006 &   0.400 $\pm$   0.012  &   1.134  $\pm$ 0.186  &  0.792  $\pm$  0.050\\
7062.25021 &  0.372  &  1.357  $\pm$  0.019    &    1.487 $\pm$   0.011   &  0.348  $\pm$  0.006 &   0.402 $\pm$   0.003  &   1.192  $\pm$ 0.118  &  0.866  $\pm$  0.114\\
7063.26832 &  0.508  &  5.586  $\pm$  0.312    &    2.111 $\pm$   0.005   &  0.414  $\pm$  0.001 &   0.452 $\pm$   0.020  &   1.16   $\pm$ 0.253  &  0.896  $\pm$  0.262\\
7064.25727 &  0.640  &  3.266  $\pm$  0.026    &    1.784 $\pm$   0.024   &  0.375  $\pm$  0.003 &   0.404 $\pm$   0.002  &   1.115  $\pm$ 0.107  &  0.828  $\pm$  0.086\\
7066.26788 &  0.908  &  1.375  $\pm$  0.005    &    1.398 $\pm$   0.012   &  0.350  $\pm$  0.001 &   0.385 $\pm$   0.002  &   1.063  $\pm$ 0.049  &  0.782  $\pm$  0.049\\
7068.27980 &  0.176  &  1.992  $\pm$  0.019    &    1.672 $\pm$   0.002   &  0.344  $\pm$  0.002 &   0.406 $\pm$   0.002  &   1.138  $\pm$ 0.23   &  0.804  $\pm$  0.218\\
7069.22819 &  0.303  &  3.114  $\pm$  0.095    &    1.896 $\pm$   0.004   &  0.401  $\pm$  0.010 &   0.454 $\pm$   0.007  &   1.227  $\pm$ 0.146  &  0.781  $\pm$  0.149\\
\hline
\end{tabular}
\end{table*}

\begin{table*}
\tiny\tabcolsep 0.3cm \caption{Our $EW$ results for excess emissions in the $\mbox{Na~{\sc i}}$\, D$_{1}$, D$_{2}$,
H$_{\alpha}$, and $\mbox{Ca~{\sc ii}}$ IRT lines for DM UMa.}
\begin{tabular}{lccccccccccll}
\hline \multicolumn{1}{c}{HJD(2450000+,)}
&\multicolumn{1}{c}{Phase} &
\multicolumn{1}{c}{EW$_{\rm Na_{I}D_{1}}$(${\AA}$)}&
\multicolumn{1}{c}{EW$_{\rm Na_{I}D_{2}}$(${\AA}$)}&
\multicolumn{1}{c}{EW$_{\rm H_{\alpha}}$(${\AA}$)}&
\multicolumn{1}{c}{EW$_{\rm Ca_{II}8498}$(${\AA}$)}&
\multicolumn{1}{c}{EW$_{\rm Ca_{II}8542}$(${\AA}$)}&
\multicolumn{1}{c}{EW$_{\rm Ca_{II}8662}$(${\AA}$)}&
\multicolumn{1}{c}{EW$_{\rm 8542}$/EW$_{8498}$}\\
\hline
7046.35136 &  0.251  & 0.086 $\pm$ 0.003  &  0.072 $\pm$    0.010&   3.770 $\pm$   0.205 &    0.854$\pm$	0.101    &     1.269$\pm$	0.014 &  0.842	$\pm$0.037  &  1.486  $\pm$	0.063\\
7049.35378 &  0.651  &   -   $\pm$    -   &    -   $\pm$      -  &   3.554 $\pm$   0.119 &    1.207$\pm$	0.095    &     1.947$\pm$	0.166 &  	-     $\pm$ -     &  1.613  $\pm$	0.089\\
7050.32868 &  0.781  & 0.085 $\pm$ 0.004  &  0.061 $\pm$    0.008&   2.681 $\pm$   0.036 &    0.978$\pm$	0.046    &     1.257$\pm$	0.124 &  1.032	$\pm$0.050  &  1.285  $\pm$	0.090\\
7051.32824 &  0.915  & 0.089 $\pm$ 0.001  &  0.042 $\pm$    0.001&   2.858 $\pm$   0.010 &    0.689$\pm$	0.047    &     0.974$\pm$	0.145 &  1.024	$\pm$0.058  &  1.414  $\pm$	0.162\\
7057.33445 &  0.716  & 0.094 $\pm$ 0.006  &  0.053 $\pm$    0.006&   3.105 $\pm$   0.027 &    0.682$\pm$	0.031    &     0.986$\pm$	0.162 &  0.836	$\pm$0.021  &  1.446  $\pm$	0.206\\
7059.31556 &  0.980  & 0.083 $\pm$ 0.001  &  0.046 $\pm$    0.005&   2.819 $\pm$   0.082 &    0.791$\pm$	0.040    &     1.186$\pm$	0.294 &  0.996	$\pm$0.221  &  1.499  $\pm$	0.338\\
7060.36109 &  0.120  & 0.079 $\pm$ 0.007  &  0.041 $\pm$    0.004&   3.333 $\pm$   0.142 &    0.769$\pm$	0.040    &     1.199$\pm$	0.310 &  1.002	$\pm$0.195  &  1.559  $\pm$	0.370\\
7061.33654 &  0.250  & 0.107 $\pm$ 0.016  &  0.071 $\pm$    0.006&   3.189 $\pm$   0.245 &    1.037$\pm$	0.036    &     1.243$\pm$	0.420 &  1.101	$\pm$0.410  &  1.199  $\pm$	0.376\\
7062.25021 &  0.372  & 0.086 $\pm$ 0.004  &  0.067 $\pm$    0.004&   3.081 $\pm$   0.198 &    1.084$\pm$	0.080    &     1.348$\pm$	0.403 &  0.974	$\pm$0.276  &  1.244  $\pm$	0.312\\
7063.26832 &  0.508  & 0.214 $\pm$ 0.010  &  0.149 $\pm$    0.001&   7.839 $\pm$   0.040 &    0.987$\pm$	0.012    &     1.499$\pm$	0.452 &  1.168	$\pm$0.193  &  1.519  $\pm$	0.450\\
7064.25727 &  0.640  & 0.106 $\pm$ 0.001  &  0.050 $\pm$    0.003&   5.336 $\pm$   0.397 &    1.200$\pm$	0.018    &     1.289$\pm$	0.539 &  1.054	$\pm$0.298  &  1.074  $\pm$	0.449\\
7066.26788 &  0.908  & 0.072 $\pm$ 0.007  &  0.050 $\pm$    0.001&   2.914 $\pm$   0.032 &    0.971$\pm$	0.096    &     1.060$\pm$ 0.300 &  0.729	$\pm$0.030  &  1.092  $\pm$	0.218\\
7068.27980 &  0.176  & 0.100 $\pm$ 0.009  &  0.066 $\pm$    0.010&   3.673 $\pm$   0.170 &    0.973$\pm$	0.031    &     1.145$\pm$	0.332 &  1.089	$\pm$0.134  &  1.177  $\pm$	0.314\\
7069.22819 &  0.303  & 0.167 $\pm$ 0.014  &  0.126 $\pm$    0.013&   4.983 $\pm$   0.054 &    0.931$\pm$	0.128    &     1.363$\pm$	0.203 &  1.096	$\pm$0.204  &  1.464  $\pm$	0.124\\
\hline
\end{tabular}
\end{table*}

\begin{table*}
\tiny\tabcolsep 0.08cm \caption{Physical parameters and chromospheric activity properties of active systems with spectral type and rotational velocity similar to DM UMa.}
\begin{tabular}{llrrrllccl}
\hline \multicolumn{1}{l}{Name}&
\multicolumn{1}{l}{Spectral type}&
\multicolumn{1}{r}{Mag}&
\multicolumn{1}{r}{Orbital period}&
\multicolumn{1}{r}{Velocity}&
\multicolumn{1}{l}{Behavior} &
\multicolumn{1}{l}{Behavior}&
\multicolumn{1}{c}{EW(${\AA}$)}
&\multicolumn{1}{c}{$\mbox{He~{\sc i}}$ D$_{3}$}
&\multicolumn{1}{l}{Reference}\\
\hline
&&  & days &  km~$s^{-1}$ & $\mbox{Ca~{\sc ii}}$ H\&K & H$_{\alpha}$ &H$_{\alpha}$ &Emission\\
\hline
  DM UMa       &  K2-3IV               & 9.31 &     7.492      &   26.0   &Strong emission    & Strong emission            & 2.2-7.8      & Yes  &Hatzes 1995; Montes et al. 1995; Our paper\\
\hline
  DF Cam       &  K2III                & 9.09 &    12.630      &   35.0   &-                  & Emission                   &    2.01      &  -   & Frasca et al. 2006                  \\
  V492 Per     &  K1III                & 6.31 &    21.289      &   27.2   &Emission           & Emission                   &    0.02      &  -   & Strassmeier et al. 1990             \\
  YZ Men       &  K1IIIp               & 7.52 &    19.310      &   20.0   &Weak emission      & -                          &      -       &  -   & Balona 1987                         \\
  VV Lep       &  K1III+F8V            & 8.34 &    10.669      &   38.7   & -                 & -                          &      -       &  -   & Eker et al. 2008                    \\
  V344 Pup     &  K1IV-III             & 6.85 &    11.761      &   34.0   &Weak emission      & Absorption                 &      -       &  -   &Cutispoto 1998                       \\
  Sigma Gem    &  K1III                & 4.14 &    19.604      &   27.5   & -                 & Filled-in absorption       & 0.18-0.34    &  No  &Montes et al. 2000                   \\
  BD+40 2194   &  K0III                & 6.66 &    23.853      &   26.3   &Strong emission    & -                          &     -        &  -   &Griffin 1994                         \\
  IL Hya       &  K1-2III-IV+G5V-IV    & 7.20 &    12.905      &   27.1   &Strong emission    & Weak emission              &    0.63      &  No  &Montes et al. 2000                   \\
  LZ Vel       &  K4III+K1-2III        & 7.18 &                &   28.0   &Weak emission      & Emission                   &  1-2         & -    & Bopp \& Hearnshaw 1983               \\
  DR Oct       &  G3-K0IV              & 8.60 &     5.574      &   21.0   &-                  & -                          &    -         & -    & Pasquini \& Lindgren 1994           \\

  HU Vir       &  K2III                & 8.55 &    10.388      &   25.0   &Strong emission    & Emission                   & 1.31-2.47    &      &Montes et al. 2000                   \\
  OX Ser       &  K0III                & 7.15 &    14.368      &   32.2   &-                  & Filled-in absorption       &     -        &  -   &Strassmeier et al. 2000              \\
  V2253 Oph    &  K0III                & 8.20 &   314.470      &   28.8   &Strong emission    & -                          &    -         &  -   &Strassmeier et al. 1994              \\
  2E1848.1+3305&  K0III-IV             &10.00 &     2.300      &   24.0   &Emission           & Emission                   &     -        &   -  &Takalo \& Nousek 1988                \\
  V1764 Cyg    &  K1III+F              & 7.69 &    40.142      &   30.0   &Emission           & Emission                   &  0.08        &  -   & Frasca \& Catalano 1994             \\
  1E1937.0+3027&  K0III-IV             &10.00 &     9.527      &   35.0   &-                  & Excess emission            &    -         & -    & Takalo \& Nousek 1988               \\
  V4091 Sgr    &  K0III                & 8.38 &    16.887      &   21.6   &Strong emission    & Absorption                 &    -         &  -   & Fekel et al. 1986                   \\
  AT Cap       &  K2III+F?             & 8.87 &    23.206      &   26.1   &Emission           & Emission                   &-             &-     & Collier et al. 1982                  \\
  HK Lac       &  K0III+F1V            & 6.66 &    24.428      &   21.1   &Emission           & Emission                   &   0.63 - 1.50& No   &Biazzo et al. 2006; Montes et al. 1997\\
  V350 Lac     &  K2IV-III             & 6.31 &    17.755      &   34.4   &Modern emission    & Absorption                 &     0.278    &  -   &Montes et al. 1995c; Fern\'{a}ndez-Figueroa et al.1994\\
  IM Peg       &  K2III+K0V            & 5.55 &    24.649      &   26.5   &Emission           & Excess emission            &-0.06 - 0.97  & No   &Biazzo et al. 2006;                   \\
\hline
\end{tabular}
\end{table*}

\clearpage



\begin{figure}
\centering
\subfigure[$\mbox{Na~{\sc i}}$ D$_{1}$, D$_{2}$]{
\includegraphics[width=16.0cm,height=11.8cm]{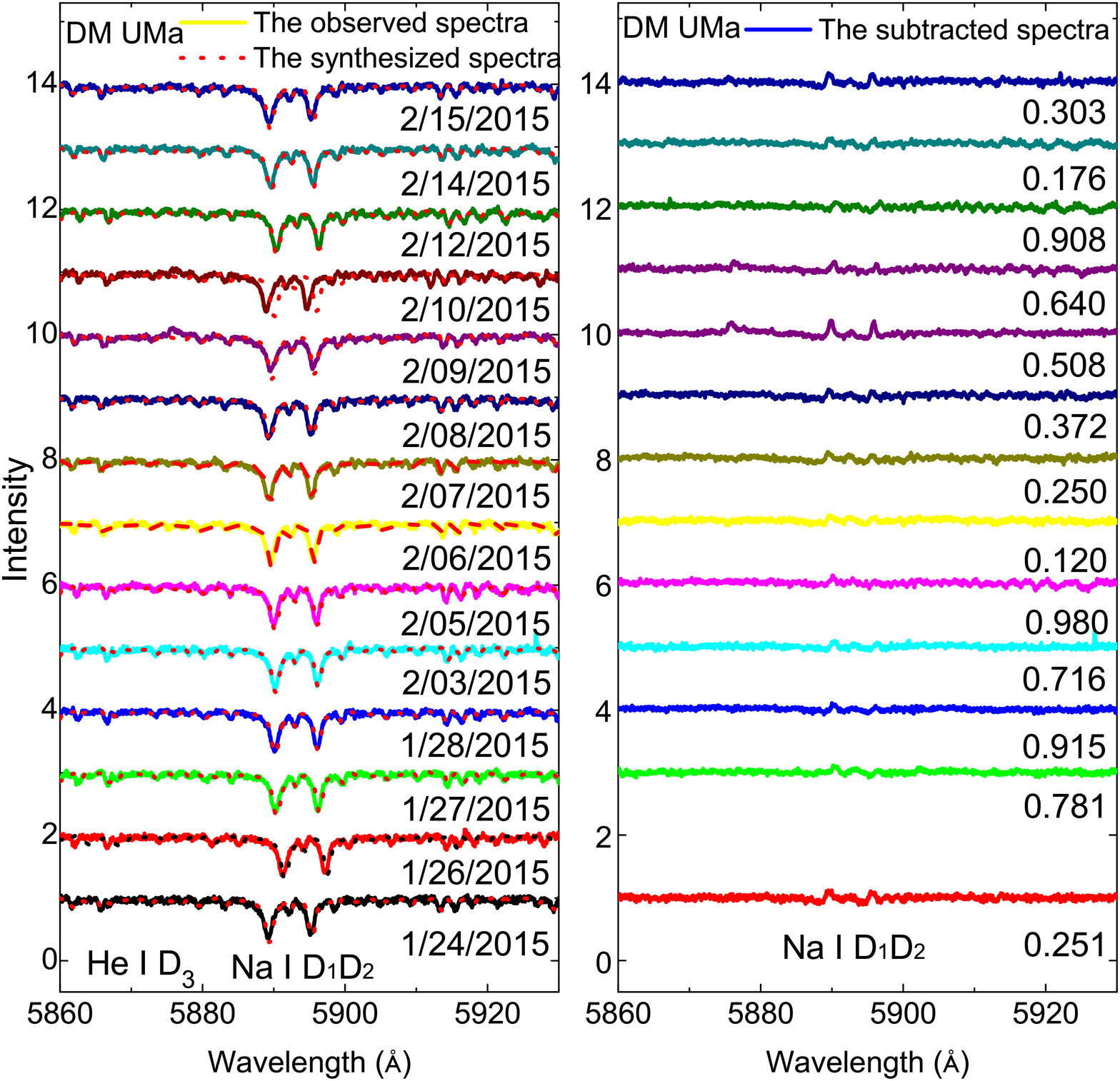}}
\hspace{1in}
\subfigure[H$_{\alpha}$]{
\includegraphics[width=16.0cm,height=11.8cm]{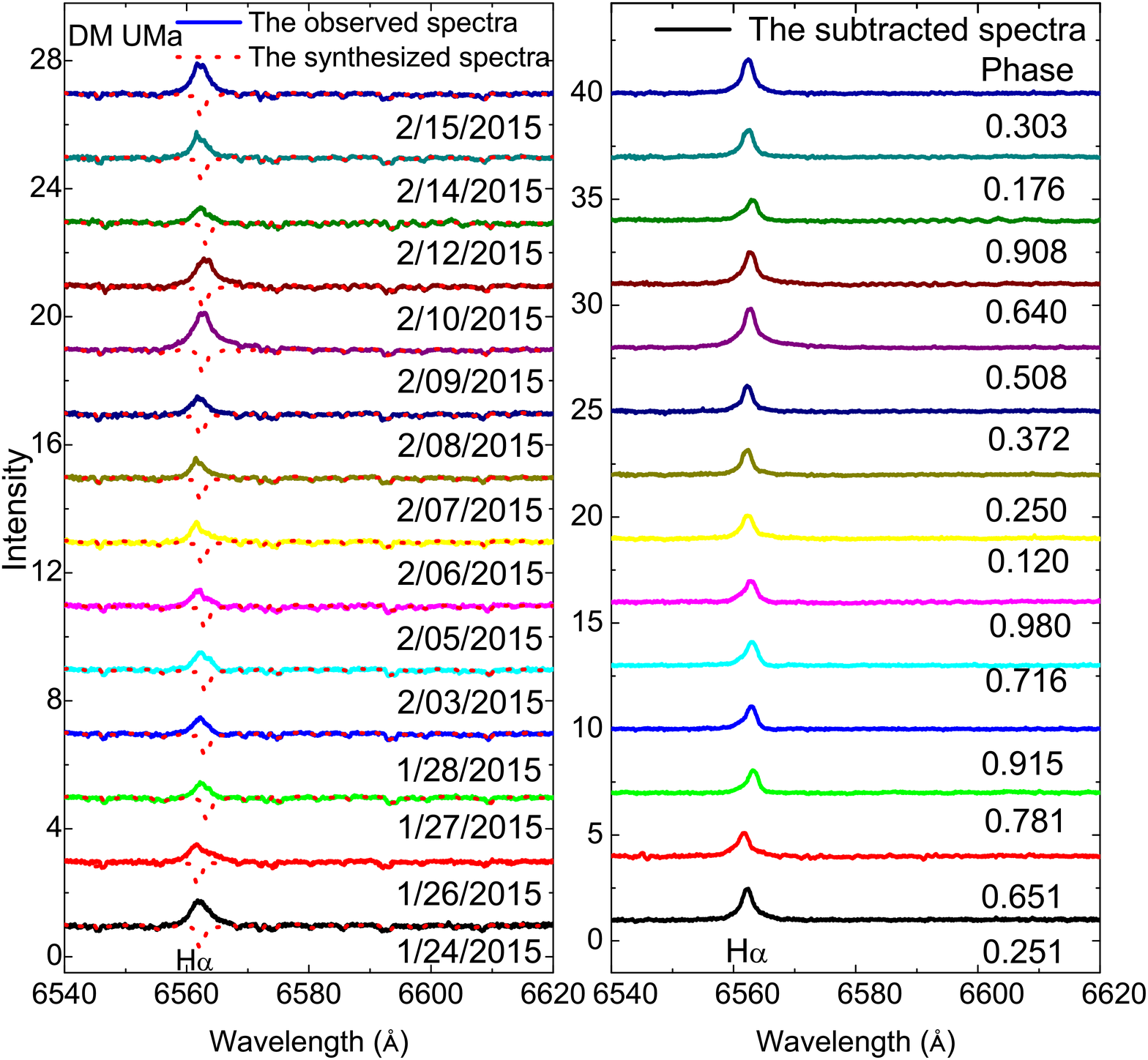}}
\end{figure}
\begin{figure}
\centering
\subfigure[$\mbox{Ca~{\sc ii}}$ IRT 8542]{
\includegraphics[width=16.0cm,height=11.8cm]{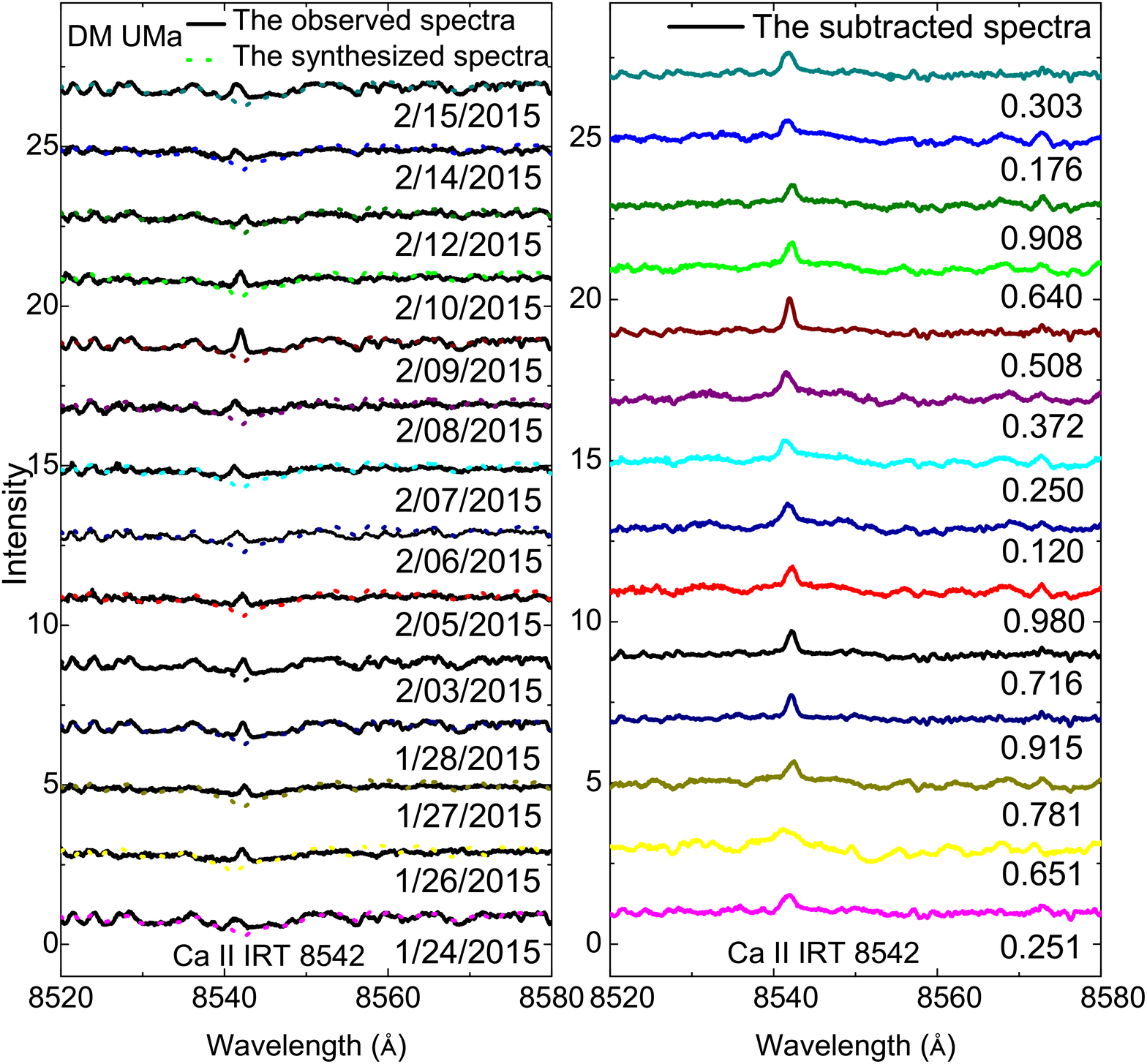}}
\hspace{1in}

\caption{Observed, synthesized, and subtracted spectra of DM UMa for several chromospheric active indicators of the $\mbox{He~{\sc i}}$ D$_{3}$, $\mbox{Na~{\sc i}}$ D$_{1}$, D$_{2}$, H$_{\alpha}$ and $\mbox{Ca~{\sc ii}}$ IRT 8542 lines. The y-axis is the spectral intensity normalized by the source flux density and the intensities are shifted on the y-axes for clarity.}\label{fig:1}
\end{figure}

 \begin{figure}
 \centering
\includegraphics[width=8.205cm,height=6.7cm]{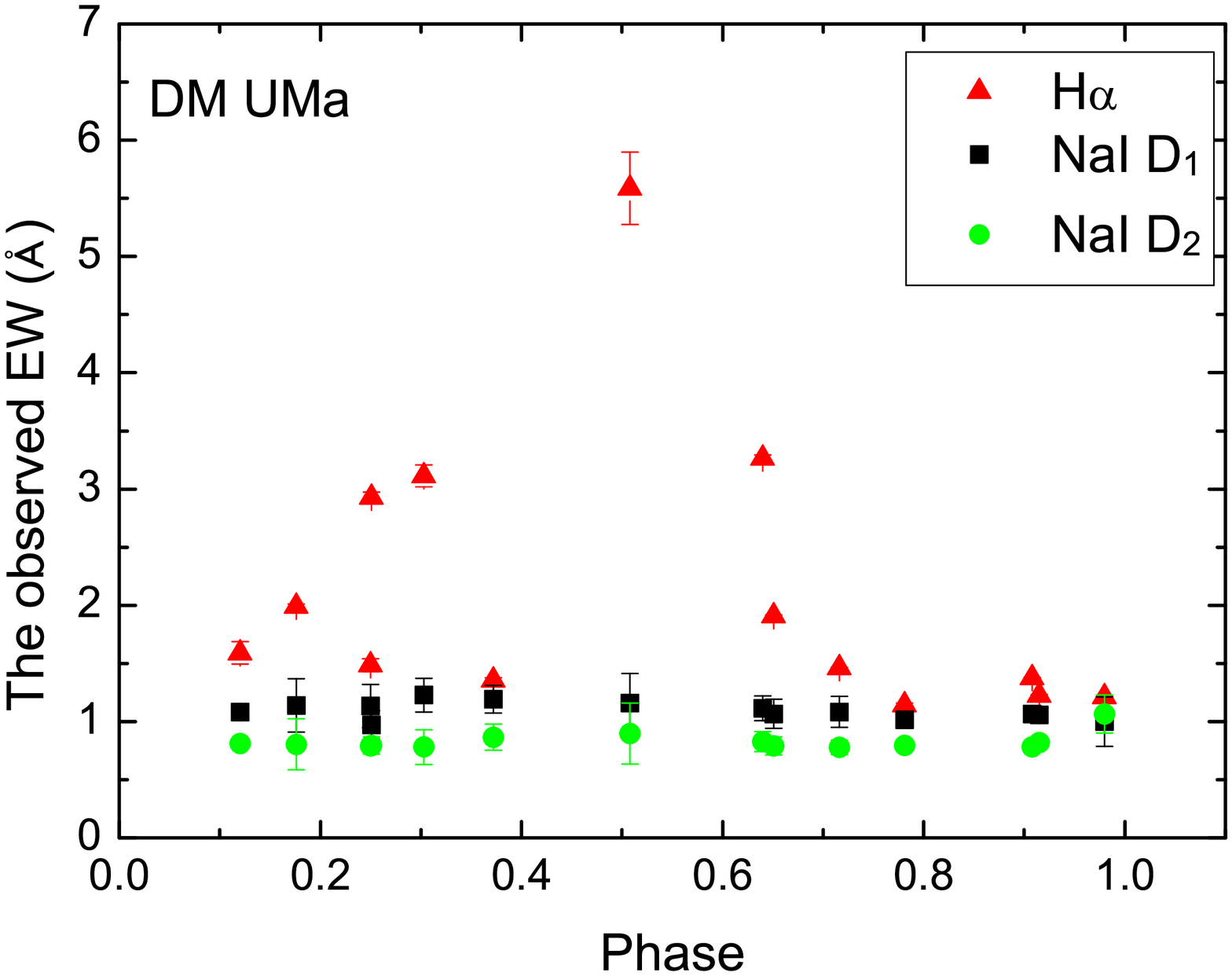}
\includegraphics[width=8.205cm,height=6.7cm]{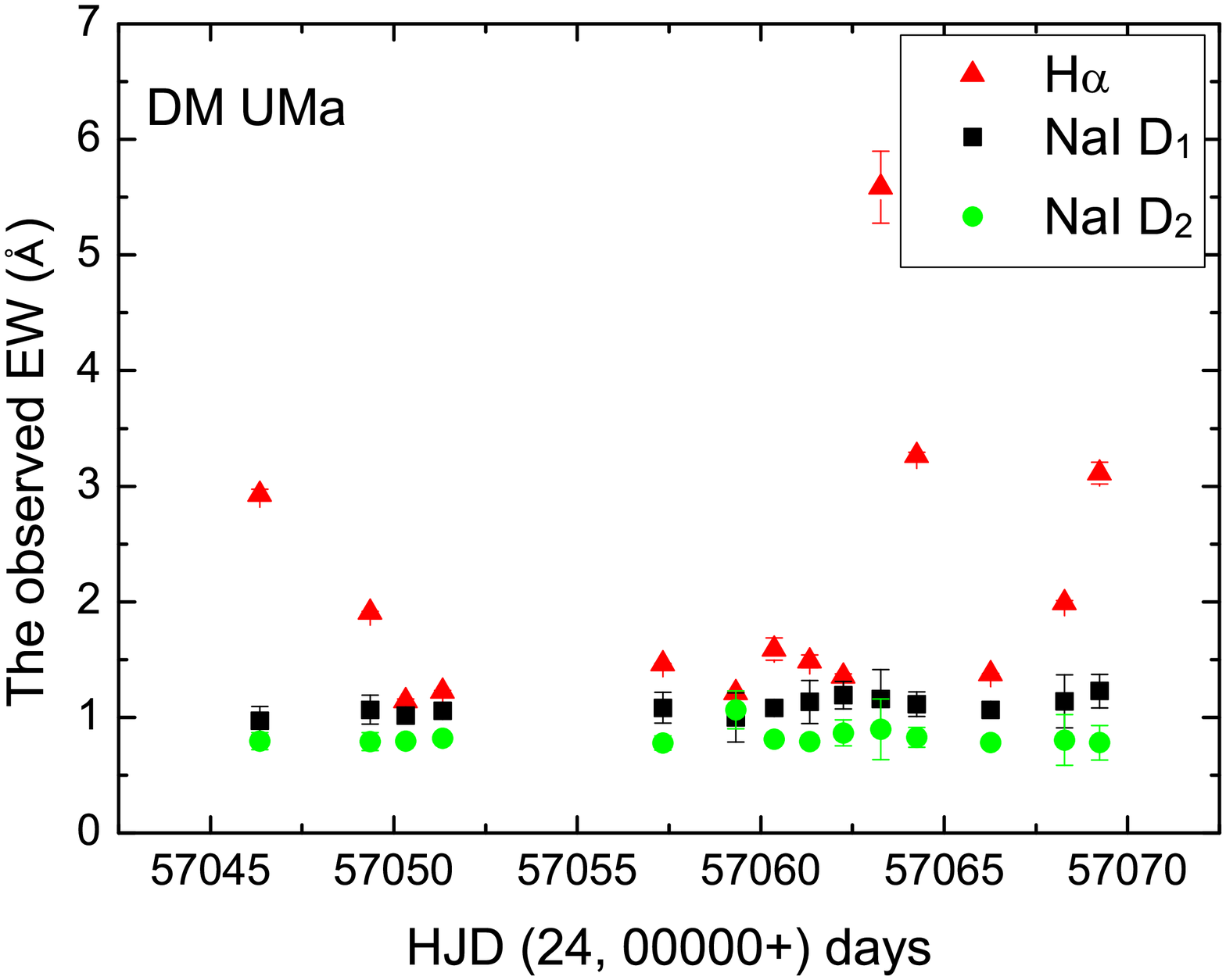}
\includegraphics[width=8.205cm,height=6.7cm]{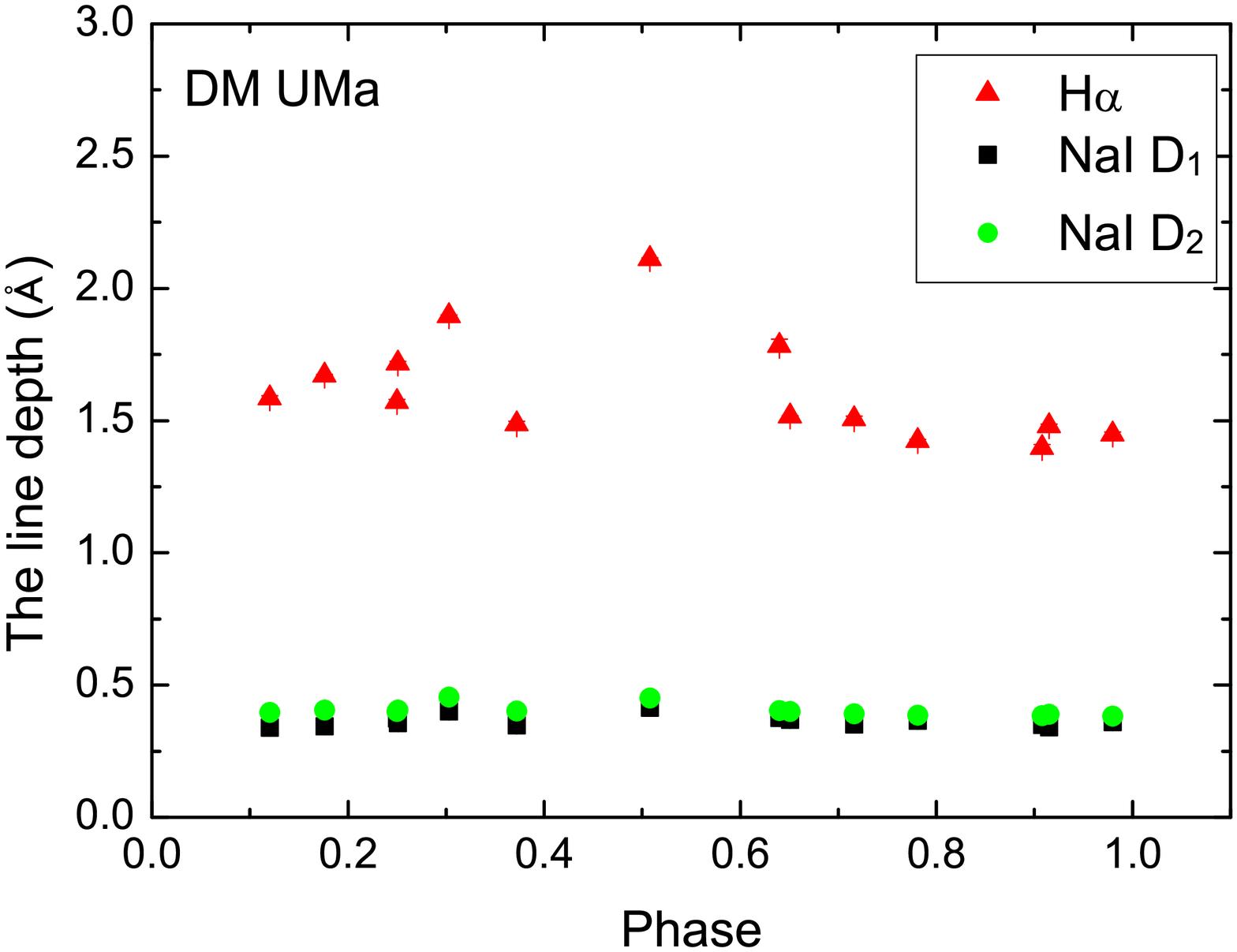}
\includegraphics[width=8.205cm,height=6.7cm]{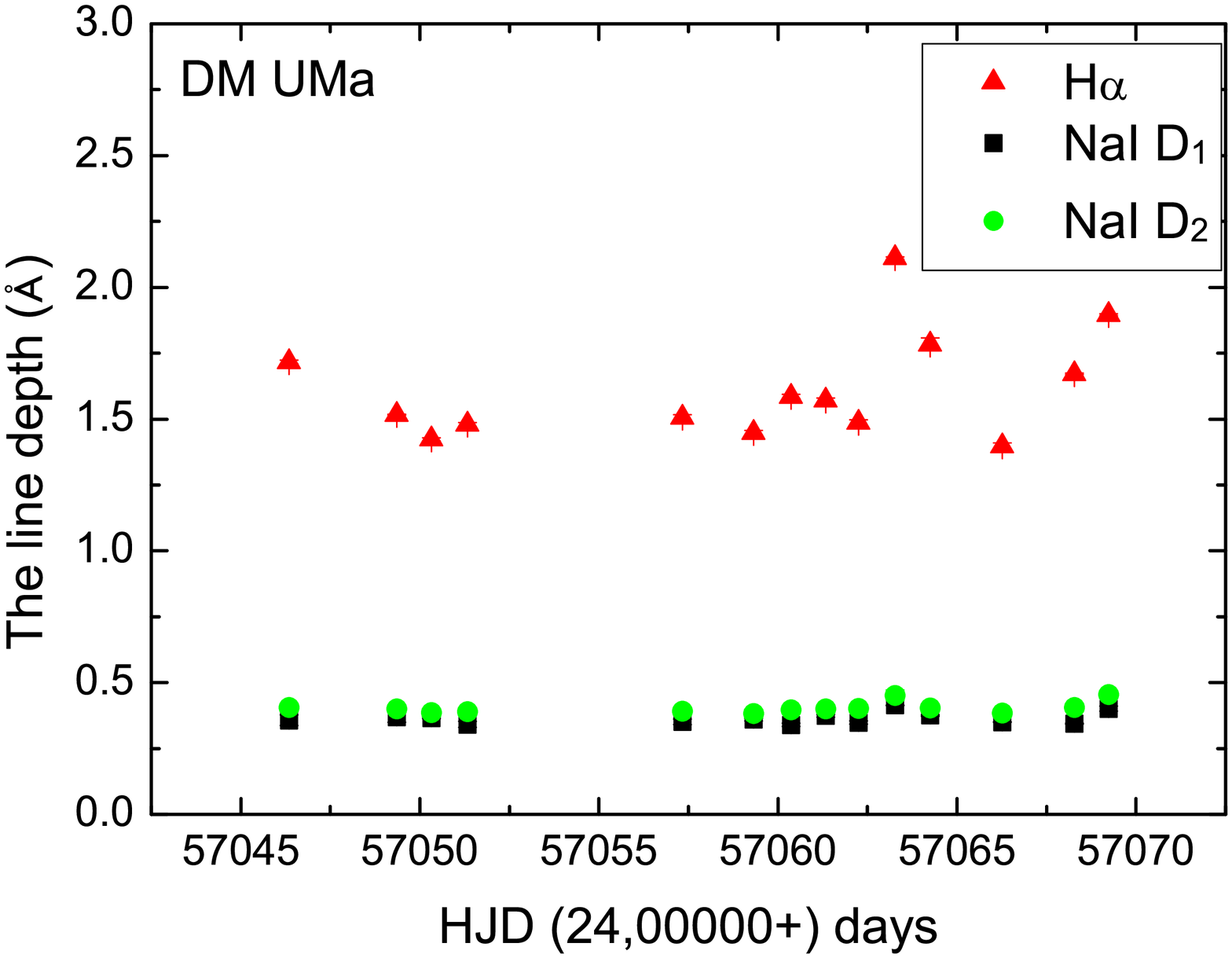}
\includegraphics[width=8.205cm,height=6.7cm]{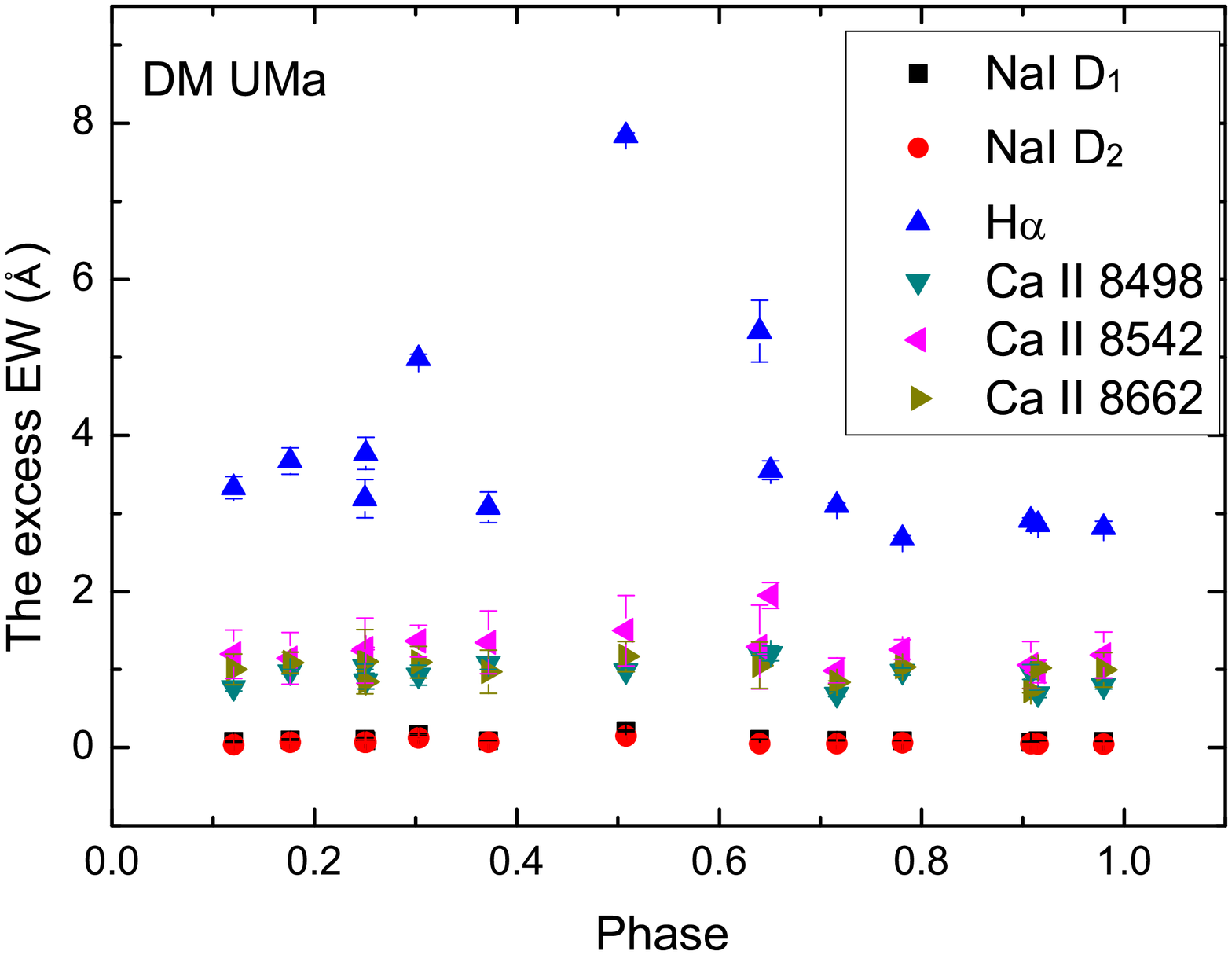}
\includegraphics[width=8.205cm,height=6.7cm]{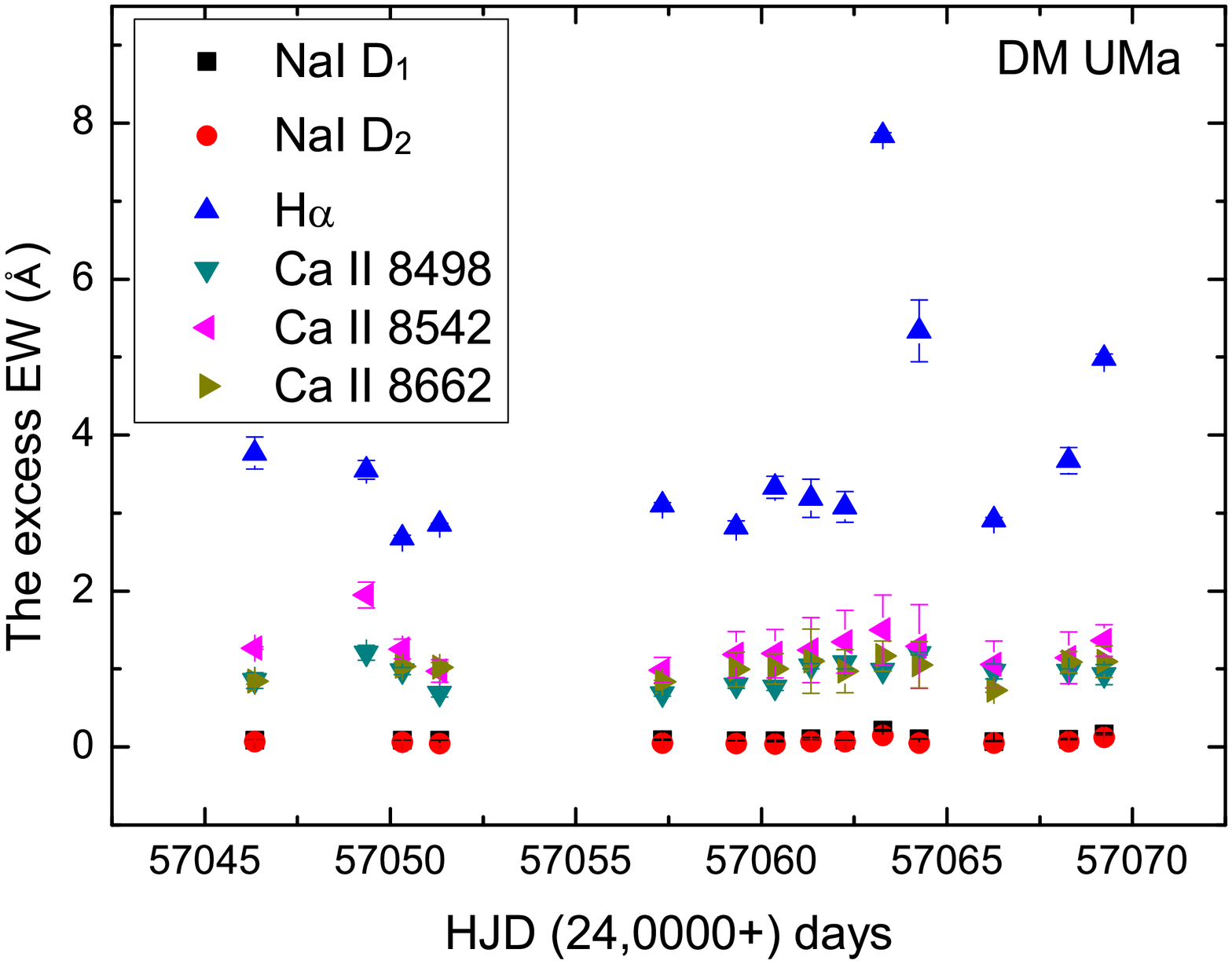}
\vspace{-0.5cm}
 \caption{$EW$s of observed emissions and line depths, and excess emissions of the chromospheric indicators of DM UMa vs. orbital phase or HJD for the $\mbox{Na~{\sc i}}$ D$_{1}$, D$_{2}$, H$_{\alpha}$ and $\mbox{Ca~{\sc ii}}$ IRT lines.}
\end{figure}

\begin{figure}
\centering
\includegraphics[width=14cm,height=10cm]{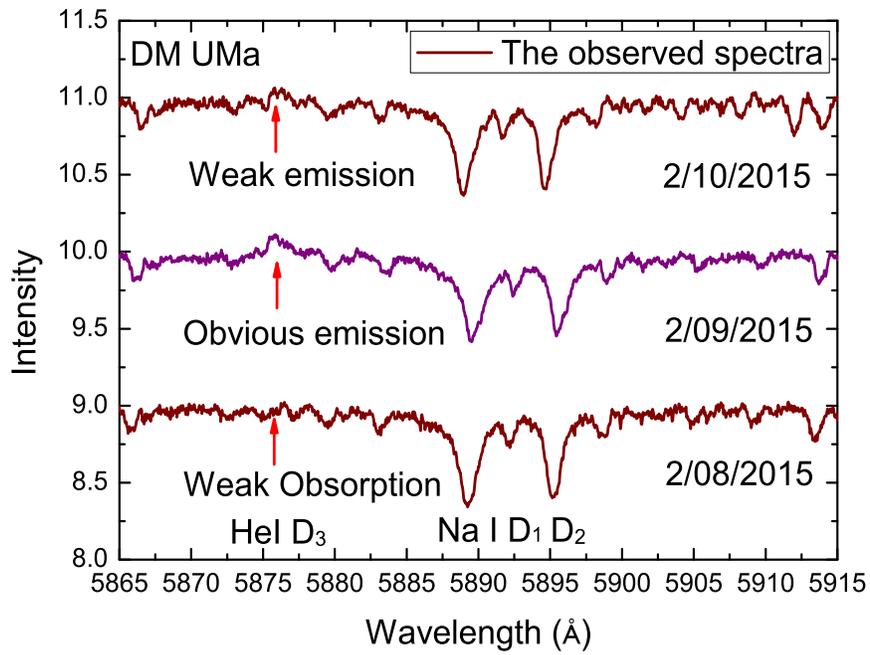}
\caption{Observed spectra of $\mbox{He~{\sc i}}$ D$_{3}$ and $\mbox{Na~{\sc i}}$ D$_{1}$, D$_{2}$ in three consecutive days. Flare events were detected with clear emission at the $\mbox{He~{\sc i}}$ D$_{3}$ line on Feb. 09, 2015 and weak emission on Feb. 10, 2015. The y-axis is the spectral intensity normalized by the source flux density and the intensities are shifted on the y-axes for clarity.}
\end{figure}


 \begin{figure}
 \centering
\includegraphics[width=12cm,height=9cm]{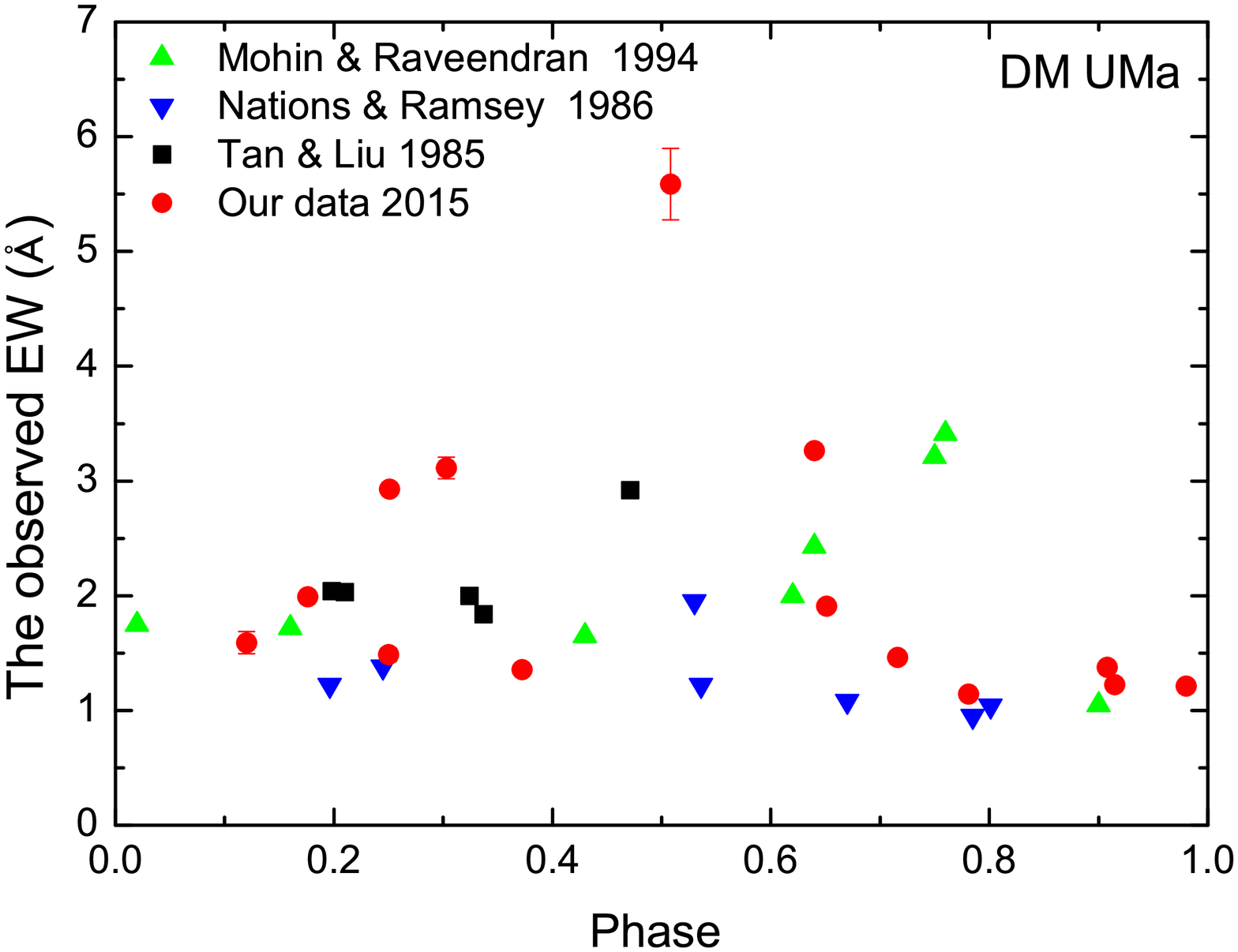}
\includegraphics[width=12cm,height=9cm]{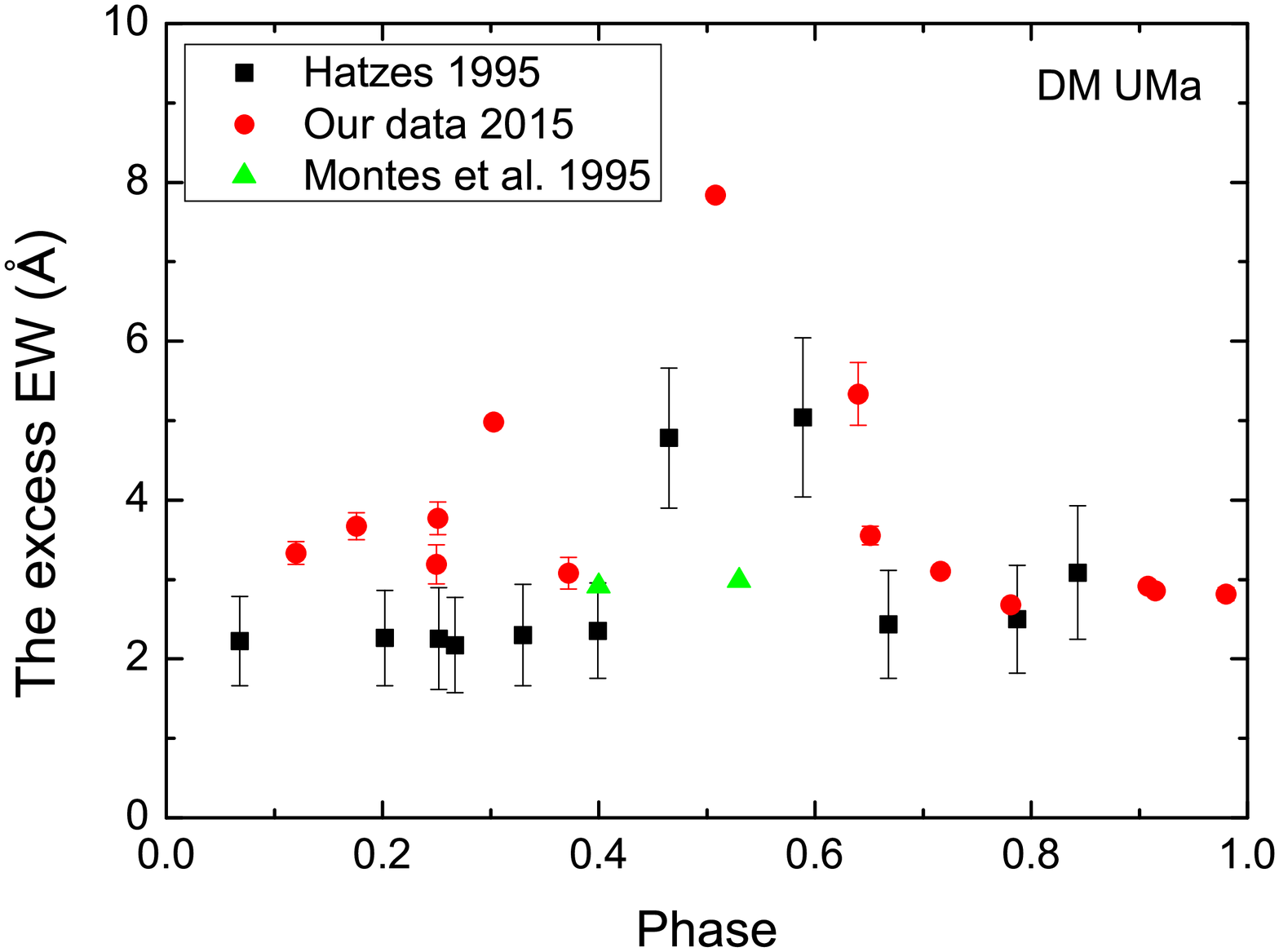}
\vspace{-0.5cm}
\caption{The Observed $EW$s and chromospheric excess emission $EW$s of DM UMa vs. orbital phase for the H$_{\alpha}$ line.}
\end{figure}
\end{document}